\documentstyle[PASJadd,epsbox]{PASJ95}
\newcommand{\vol}{52}
\newcommand{\no}{1}
\newcommand{\lett}{}
\newcommand{\spage}{\bf ??}
\newcommand{\rdate}{1999 November 26}
\newcommand{\adate}{1999 December 22}

\begin{document}

\title{Color--Magnitude Sequence in the Clusters at $z$ $\sim$ 1.2 \\
near the Radio Galaxy 3C 324} 

\author{Masaru {\sc Kajisawa}, Toru {\sc Yamada}, Ichi {\sc Tanaka} \\
{\it Astronomical Institute, Tohoku University, Aoba-ku, Sendai, Miyagi 980-8578} \\
{\it E-mail(MK): kajisawa@astr.tohoku.ac.jp } \\
 \\
Toshinori {\sc Maihara}, Fumihide {\sc Iwamuro}, Hiroshi {\sc Terada},
Miwa {\sc Goto}, \\
Kentaro {\sc Motohara}, 
Hirohisa {\sc Tanabe}, Tomoyuki {\sc Taguchi}, Ryuji {\sc Hata} \\
{\it Department of Physics, Faculty of Science, Kyoto University,}
{\it Sakyo-ku, Kyoto 606-8502} \\
 \\
Masanori {\sc Iye}, Masatoshi {\sc Imanishi}, Yoshihiro {\sc Chikada},
Michitoshi {\sc Yoshida} \\
{\it National Astronomical Observatory, 2-21-1 Osawa, Mitaka, Tokyo
 181-8588} \\
 \\
Chris {\sc Simpson}, Toshiyuki {\sc Sasaki}, George {\sc Kosugi},
Tomonori {\sc Usuda}\\
and\\
Kazuhiro {\sc Sekiguchi} \\
{\it Subaru Telescope, National Astronomical Observatory of Japan,} \\
{\it  650 North Aohoku Place, Hilo, HI 96720, U.S.A. }\\
\vspace{0.5cm}
 }

\abst{
 We have investigated the optical and near-infrared colors of
$K'$-selected galaxies in clusters at $z \sim 1.2$ near to the radio
galaxy 3C 324 using images obtained with the Subaru telescope and
archival HST data. The distribution of colors of
the galaxies in the cluster region is found to be fairly broad, and 
it may imply  
significant scatter in their star-formation histories,
although the effect of contamination of field galaxies is uncertain.
The red sequence of galaxies whose $R-K$ colors are consistent with
passive evolution models for old galaxies is 
found to be truncated at $K' \sim 20$ mag, and there are few fainter
galaxies with similar red colors in the cluster region.
We find that the bulge-dominated galaxies selected by quantitative
morphological classification form a broad sequence in the
color--magnitude diagram, whose slope is much steeper than that
expected from metallicity variations within a passively evolving
coeval galaxy population. We argue that the observed color--magnitude
sequence can be explained by metallicity and age variations, and the
fainter galaxies with $K' > 20$ mag may be 1--2 Gyr younger than the
brighter galaxies. Some spatial segregation of the
color and $K'$-band luminosity is seen in the sky distribution; the
redder and the brighter objects tend to be located near 3C 324.}
\kword{galaxies: clusters of --- galaxies: evolution --- galaxies: formation}

\maketitle
\thispagestyle{headings}

%\vspace{0.5cm}
%\clearpage
\section{Introduction}
  By tracing the average properties of galaxies in rich clusters and
their progenitors from low to high redshift, we can study how galaxies
in similar environments evolve with cosmic time, and obtain insights
into the phenomenon of galaxy formation. In nearby and
intermediate-redshift clusters, the majority of bright galaxies are
known to form a tight and conspicuous red sequence in the color--magnitude
diagram (e.g., Visvanathan, Sandage 1977; Bower et al. 1992). This
color--magnitude (hereafter C--M) relation seen in nearby clusters
extends over a range of at least 4-magnitudes (Garilli et al. 1996)
and possibly 5--6 mag (e.g., De Propris et al. 1998) from the brightest
cluster galaxies; the slope of the relation does not vary from
cluster to cluster.  Bower et al. (1992) showed that the small
dispersion of the C--M relation for early-type galaxies in the Coma and
Virgo clusters places a strong constraint on their age or coevality by
assuming a simple star-formation history; they found that the
early-type galaxies in these clusters are older than $\sim$ 10 Gyr, if
an age spread of $\sim 1$ Gyr is assumed (see Bower et al. 1998 for
more general discussion). Arag\'on-Salamanca et al. (1993) then found
a systematic evolution in the colors of the reddest cluster members
which they interpreted as being passive evolution of old galaxies formed at
$z \gtsim 2$. Ellis et al. (1997) and Stanford et al. (1998)
investigated {\it morphologically selected} early-type galaxies in
intermediate-redshift clusters, and found that the well-defined C--M
sequence with small scatter holds there. Gladders et al. (1998) also
found that the evolution of the slope of the C--M sequence is
consistent with the predictions of passive evolution. These
conclusions from studies of the C--M relation are complemented by those
from Fundamental Plane analysis (e.g., van Dokkum et al. 1996,
1998). The C--M relation is more favorably interpreted as being a metallicity
sequence, rather than an age sequence, since otherwise the slope should
be much steeper than observed even at an intermediate redshift (Kodama,
Arimoto 1997), although some age difference between bright and
faint galaxies has been pointed out by some authors based on the
observed differences in the absorption-line strength (e.g., Terlevich et
al. 1999).

 On the other hand, there are many observational indications that the
star-formation activity in more distant clusters is higher than that
in local clusters. It is known that the fraction of blue galaxies in
clusters increases rapidly with the redshift at intermediate-redshift
(Butcher, Oemler 1978, 1984; Rakos, Schombert 1995). Rakos and
Schombert showed that the fraction of blue galaxies becomes $\sim$
80\% at $z =$ 0.9. Through spectroscopic studies of
intermediate-redshift clusters, it is also known that a significant
fraction of galaxies have emission-lines and/or post-starburst
signatures (e.g., Dressler, Gunn 1992; Postman et al. 1998;
Dressler et al. 1999; Poggianti et al. 1999). Postman et al. (1998)
investigated two clusters at $z \sim$ 0.9 spectroscopically, and found
that about half of the galaxies show high levels of star formation
activity. It is expected from these results that a more active evolution
of galaxies may be observed in clusters at higher redshift.

 Recently, a number of clusters and cluster candidates at $z$ $\gtsim$
1 have been discovered (Dickinson 1995; Yamada et al. 1997; Stanford
et al. 1997; Hall, Green 1998; Ben\'\i tez et al. 1999; Tanaka et
al. 1999; Rosati et al. 1999). By observing high-redshift clusters,
any differences in the star-formation history can be seen more clearly.
Tanaka et al. (1999) investigated the colors of galaxies in the
cluster 1335.8+2820 at $z \sim$ 1.1, and showed that the optical and
near-infrared color distributions of the galaxies in this cluster are
wide, which may be attributed to variations in the star-formation
activity. They found that the fraction of galaxies which show some
UV excess over the expected colors for passive evolution is more than
75\% in this cluster.  Stanford et al. (1997) investigated the
NIR-selected cluster CIG J0848+4453 at $z =$ 1.27, and also showed that
those galaxies whose $RJK$ colors are consistent with passively evolving
elliptical models have relatively blue $B-K$ colors, which indicates
that some star-formation activity is occurring in the old galaxies in
the cluster core. It is therefore interesting to expand the detailed
color analysis to other clusters at $z$ $\gtsim$ 1, and to see whether
these results apply generally to high-redshift clusters.

 In this paper, we consider the optical and near-infrared (NIR)
colors of $K'$-selected galaxies in those clusters at $z \sim$ 1.2 near
to the radio galaxy 3C 324 (Dickinson 1997a,b; Kajisawa, Yamada 1999;
Kajisawa et al. 2000, hereafter as Paper I) using images
obtained with the Subaru telescope
and archival data from the Hubble Space Telescope (HST).

 An excess of galaxy surface density in this region was recognized by
Kristian et al. (1974) and by Spinrad and Djorgovski (1984), and firmly
identified by Dickinson (1995). In Dickinson's (1997b) spectroscopic
study, the surface-density excess was revealed to be due to the
superposition of two clusters or rich groups at $z=1.15$ and
$z=1.21$.  Extended X-ray emission with a luminosity comparable to
that of the Coma cluster has been detected in the direction of 3C 324
(Dickinson 1997a), which suggests that at least one of the two systems
is a fairly collapsed massive system. Smail and Dickinson (1995)
detected a weak shear pattern in the field that may be produced by a
cluster. In the following discussion, we do not distinguish the two
systems at $z \sim 1.2$, since a detailed redshift distribution of
the galaxies or the relative population of them is still not
available.  We thus discuss the average properties of the two
clusters. Since their redshifts are close, it has little effect on our
discussion about the luminosity and color distributions.

 Using a $K'$-selected sample provides such advantages as (i)
the selection bias at large redshift is small, and thus a direct comparison
with lower-redshift clusters is possible (Arag\'on-Salamanca et
al. 1993) and (ii) the sample selection is less affected by the
star-formation activity. Furthermore, the rest-frame mid-UV color provided
by the HST data enables us to detect even a small amount of recent
star-formation (e.g., Ellis et al. 1997; Smail et al. 1998).  We
describe the observations and data reduction briefly in section 2. In
section 3, we present the color-magnitude and two-color diagrams. Our
conclusions and discussions are given in the last section.

 In a previous Paper I,
we studied the $K$-band luminosity distribution of the galaxies in
these clusters near to 3C 324 using the same $K'$-band data, which
is closely related to the results presented here. In summary,
Paper I showed that the shape of the bright end ($K \ltsim 20$) of
the luminosity function is similar to those of lower-redshift clusters;
also the measured value of the characteristic magnitude, $K^*$, follows
the evolutionary trend of intermediate-redshift clusters, which is
consistent with passive evolution models with $z_{\rm form} \gtsim 2$. At
the faint end, however, the excess galaxy surface density within 40''
 of 3C 324 decreases abruptly at $K \sim 20$, which may indicate
a strong luminosity segregation with distance from 3C 324 or a real
deficit in the faint galaxy population in the clusters. Kajisawa and
Yamada (1999) also presented results from a study of the colors and
morphologies of an {\it optically selected} sample of cluster galaxies
using the same HST data.

\section{Observations and Data Reduction}
\subsection{$K'$-Band Image}

 $K'$-band imaging of the 3C 324 field was carried out with the Cooled
Infrared Spectrograph and Camera for OHS (CISCO, Motohara et al. 1998)
mounted on the Subaru Telescope on 1999 April 1 and 2 (UT),
during the telescope commissioning period. The detector is a 1024
$\times$ 1024 HAWAII HgCdTe array with a pixel scale of 0.\hspace{-2pt}''116,
which provides a field of view of $\sim$ 2' $\times$ 2'. The presently 
images studied are common to those analyzed in
Paper I. See Paper I for details concerning the data reduction.
The seeing size of the resultant frame is 0.\hspace{-2pt}''8 arcsec as measured from
the FWHM of the stellar images.

 A flux calibration was performed by observing one the UKIRT Faint
Standard FS 27 immediately after the 3C 324 field at a similar zenith
distance. We used the empirical relation $K' - K = 0.2 (H - K)$
(Wainscoat, Cowie 1992) in order to derive the $K'$ mag of
FS 27 from its $H$ and $K$ mag.

 Source detection was performed with the SExtractor image analysis
package (Bertin, Arnouts 1996). We used a detection threshold of
$\mu_{K'} =$ 22.4 mag arcsec$^{-2}$ over 20 connected pixels. We
removed from the final catalog those bright objects with $K' <$ 18, which
appeared unresolved. Although we did not make any star/galaxy separation at $K'
> 18$, the contamination by stars in the catalogue is small at $K'
> 18$ (at most $\sim$ 5 -- 10\%; De Propris et al. 1999). In total,
146 sources were cataloged.  We used a 3'' diameter aperture to
measure the $K'$ mag of the objects and a 1.\hspace{-2pt}''6 one to
derive the colors.

\subsection{Optical Images}

 To investigate the colors of the $K'$-selected galaxies, we analyzed
the archival HST/WFPC2 F702W and F450W images of the 3C 324 field
(PI: M.Dickinson; PIDs 5465 and 6553, respectively). The total
exposure times were 64800 sec for the F702W image and 17300 sec for
the F450W image.

 After combining the separate exposures with cosmic-ray rejection, we
aligned the F702W and the F450W images to the $K'$ image using
point sources common to the frames. The optical images were
 then convolved with a Gaussian kernel to match the FWHM of the stellar
images in the $K'$ frame. Photometry of each $K'$-selected object
was performed using apertures with identical positions and size
(1.\hspace{-2pt}''6 diameter) regarding the $K'$ image in each optical
frame. We used the STMAG system for these optical magnitudes, and
hereafter refer them as $R_{\rm F702W}$ and $B_{\rm F450W}$ (STMAG
is defined as $m_{\rm ST} = -21.10-2.5 {\rm log} f_\lambda$ where
$f_\lambda$ is expressed in erg cm$^{-2}$ s$^{-1}$ \AA $^{-1}$; 
HST Data Handbook Vol.1). Due to the limitation of the WFPC2
field of view (figure 10), the number of $K'$-selected galaxies for
which $R_{\rm F702W}$ and $B_{\rm F450W}$ mag were measured is 139.

\section{Results}

\subsection{Color--Magnitude Diagram}

 Figure 1 shows the C--M diagram ($K'$ vs $R_{\rm F702W}-K'$) of the
detected galaxies. We have divided the observed frame into the
`cluster' region (within 40'' of 3C 324) and the `outer' region
(the remainder of the frame); note that a clear excess of galaxies
with $K$ $\ltsim$ 20 mag in the `cluster' region has been recognized
at $\sim 10 \sigma$ significance (Dickinson 1995; Paper I). 
The areas of the `cluster' and the `outer' regions are
1.193 arcmin$^{2}$ 
and 1.877 arcmin$^{2}$, respectively, and the galaxies in each region
are plotted with different symbols in the figure. The error bars are
based on the 1$\sigma$ background fluctuation within a 1.\hspace{-2pt}''6
aperture. The dotted line represents the $\sim$ 70\% completeness
level of $K'$ = 21.5 (Paper I). The dashed line shows $R_{\rm F702W}
= $ 28 mag, which corresponds to the $\sim$ 3$\sigma$ detection limit
in this band. Most of the $K'$-selected galaxies, except for a few
extremely red objects, have $R_{\rm F702W}$ magnitudes brighter than this
limit.

 In figure 1, the red envelope of the galaxies with $R_{F702W} - K'
\sim$ 6 can be seen, and those objects with $K' \ltsim$ 21 mag and
$R_{\rm F702W} - K' >$ 4.5 are dominated by the galaxies in the `cluster'
region, despite the area of the outer region being 1.6-times larger
than that of the cluster region. In the magnitude range 17 $< K' <$
21, there are 6$\pm$2, 8$\pm$1, and 5$\pm$3 cluster galaxies with
$R_{\rm F702W}-K' >$ 5.5, 4.5 $< R_{\rm F702W}-K' <$ 5.5, and $R_{\rm
F702W}-K' <$
4.5, respectively, after applying a field correction using the `outer'
region. The quoted uncertainties are based on Poisson statistics.
Thus, about one-third of the cluster galaxies with $K' <$ 21 mag have
$R_{\rm F702W}-K' >$ 5.5 and their stellar populations are expected to be
fairly old ($\gtsim$ 2 Gyr); a more detailed comparison with the
galaxy evolutionary models is presented below.

 We confirm the existence of the `red finger', a sequence of galaxies
with $R_{\rm F702W}-K' \sim$ 6 and $K' \sim$ 18--19 mag, which was
identified by Dickinson (1995). The observed colors of these galaxies
are consistent with Dickinson's result if we consider $R_{\rm F702W} - R$
and $K' - K$ color differences. Kodama et al. (1998) derived the slope
and zero point of this `red finger' using the data of Dickinson
(1995), and showed that the sequence is consistent with a metallicity
sequence at $z \sim 1.2$ for passively evolving galaxies. This red
sequence of galaxies in the cluster region, however, seems to be
truncated at $K'$ $\sim$ 20. Although there are a few red objects in the
outer region, there is no such object in the cluster region fainter
than this magnitude. This is {\it not} due to incompleteness; at $K'
\sim 21$ the catalog is still $\sim 90$\% complete (Paper I).
To emphasize this point, figure 2 shows the number counts of the
`red sequence' galaxies with 5.7 $< R_{\rm F702W} - K' <$ 6.3, which
corresponds to the color range adopted in Kodama et al. (1998),
together with the detection completeness in the $K'$-band evaluated in
Paper I.

 Another important aspect of the C--M diagram in figure 1 is that the
color distribution of galaxies in the cluster region is fairly wide:
in addition to the `red sequence' galaxies, there are many galaxies
with bluer $R_{\rm F702W} - K'$ colors in the cluster region. There seems
to exist an even more significant surface density excess within the
`cluster' region in the range 4.5 $< R_{\rm F702W} - K' <$ 5.5 ($\sim 8
\pm 1$ cluster galaxies) compared with those in the $R_{\rm F702W} - K' >
5.5$ range ($\sim 6 \pm 2$ cluster galaxies). At $R_{\rm F702W} - K' <$
4.5, a marginal excess of cluster-region galaxies can also be seen
(5$\pm$3 cluster galaxies). The galaxies with bluer $R_{\rm F702W} - K'$
colors tend to be fainter at $K'$. The median magnitude of the
galaxies with $4.5 < R_{\rm F702W} - K' < 5.5$ is $K' \sim 20$, while that
of the redder ones is $K' \sim 19$.

\subsection{Two Color Diagram}

 The wide spread in the optical-NIR colors may be due to age differences
and/or ongoing star formation as well as to contamination by
foreground/background galaxies. To investigate the causes of the large
scatter in the C--M diagram, we show a $B_{\rm F450W}-R_{\rm F702W}$ vs
$R_{\rm F702W}-K'$ two-color diagram for the galaxies with $K' < 22$ in
figures 3 and 4. The shaded and open symbols in figure 3 represent the
galaxies in the `cluster' and the `outer' regions, respectively.
The arrows show that the objects are fainter than the 3 $\sigma$ upper
limit of $B_{\rm F450W}$ = 26.

 In figure 4, we add the model colors of galaxies with various
star-formation histories observed at $z =$ 1.2, using the GISSEL96 code
(Bruzual, Charlot 1993), and adopting the Salpeter initial mass
function. We examined the star formation models with a 1 Gyr single burst,
exponentially decaying star-formation rates with timescales $\tau =$
0.5 and 1 Gyr, and a constant star formation rate (SFR). We also
investigated a 1 Gyr single burst model with a metallicity of 0.2-times
the solar value, and confirmed that the differences between it and the
solar-metallicity model are seen along the `age direction', as
expected from the ``age--metallicity degeneracy''. The colors of the
models of old-population plus ongoing starburst are also
plotted; to produce these we have added the constant SFR models with
an age of 0.1 Gyr and mass fractions from 0.02\% to 0.5\% to the old
single-burst models. It can be seen that these models may be mimicked by
the exponentially decaying models. Note that these models have much
redder colors than those for spiral galaxies, which are
characterized by $\tau >$ 2 Gyr (Bruzual, Charlot 1993).

 It is found from figure 4 that the colors of the galaxies in the red
sequence, $R_{\rm F702W} - K' > 5.5$, are consistent with those of the
models whose age is 3--4 Gyr, irrespective of the existence of a small
amount of ongoing star formation.

 On the other hand, many of the galaxies with $4.5 < R_{\rm F702W} -K' <
5.5$ are more consistent with the models with a younger age, 1--3 Gyr,
and are insensitive to the amount of star-formation activity, if we assume
they are at the same redshift and neglect possible reddening by dust.
The bluer $R_{\rm F702W} - K'$ colors must be due, at least in part, to a
younger age, since the $B_{\rm F450W}-R_{\rm F702W}$ color would be $\sim$
0.5--1 mag bluer than observed if the bluer $R_{\rm F702W} - K'$ colors
were purely due to star-formation activity. In fact, the bluer
$R_{\rm F702W} - K'$ colors must be {\it predominantly} due to
a youthfulness effect for objects with $B_{\rm F450W}-R_{\rm F702W} \gtsim 0$,
which are less affected by star formation. Note that the effect of
metallicity on the age estimation is small in this range. Although there may be
effects of dust extinction, the reddening arrow (Cardelli et
al. 1989) in figure 4 is almost parallel to the lines of constant age.
There are also some galaxies in the color range $B_{\rm F450W} -
R_{\rm F702W}
\ltsim$ 0 and $R_{\rm F702W} - K' \sim$ 4, which may be consistent with
$\sim$ 1 Gyr-old galaxies at $z =$ 1.2 or galaxies with strong
star-formation, although the contamination by foreground/background
galaxies is expected to be large in this color range. From these
results, both the ages and star formation histories of the 3C 324
cluster galaxies appear to have significant scatter.

 In figure 4, we have further divided the $K'$-selected galaxies into
two samples with $K' <$ 20 (circles) and 20 $< K' <$ 22
(squares).  As inferred from figure 2, there are few `red sequence'
galaxies with $K' >$ 20 mag. On the other hand, most galaxies with
$R_{\rm F702W} - K' \sim$ 4.5 whose colors are consistent with those at $z
\sim 1.2$ have $K' >$ 20. The population with $R_{\rm F702W} - K' \sim$
4.5--5.5 is a mixture of the bright and faint samples. There may be a
trend that galaxies with younger ages or stronger star-formation
activity are fainter in the $K'$-band.  Postman et al. (1998)
investigated two clusters at $z \sim$ 0.9, and showed the existence of
a correlation between the `color age' and luminosity (their figures 18
and 19). The 3C 324 clusters may have a similar property.

 The arguments about the star-formation histories of the galaxies
presented above are based on the assumption that many of these red
galaxies are indeed at $z=1.2$. In order to further understand the
properties of
the cluster galaxies, we filtered the sample to be less
contaminated by the foreground galaxies. Most of the galaxies with
$R_{\rm F702W} - K'$ bluer than the passive evolution models at $z =$ 1.2
at a given $B_{\rm F450W} - R_{\rm F702W}$ color are considered to be in the
foreground. We tentatively divided the two-color diagram into two color
domains of `$z=1.2$' and `foreground' using as a boundary the 1 Gyr
burst model with 0.2 solar metallicity. The galaxies in the `$z=1.2$'
color domain are expected to be dominated by the cluster members,
although some of them may be dusty foreground galaxies or relatively
young galaxies at a higher redshift. As a further check, we compared the
number counts of galaxies in the `foreground' color domain located in
the `cluster' region with the general field counts derived by
averaging the number counts taken from the literature (see Paper
I) in figure 5. They are fairly consistent down to $K
\sim 21.5$, where our detection incompleteness becomes as large as
$\sim 30 \%$.

 At $K \gtsim 20$, even the total number counts of the `cluster'
region are slightly lower than the average field counts. At the same
time, there are not as many galaxies in the `cluster' region with $K >
20$ in the `$z = 1.2$' color domain, as expected for a luminosity
function with a steep faint-end slope, $\alpha \ltsim -1$. For
example, there are four galaxies in the`$z=1.2$' color domain between
$20.5 < K' < 21.5$, while the expected number is 9.1 for $\alpha=-0.9$
(Paper I). The apparent deficiency of the surface density excess
in the `cluster' region discussed in Paper I may thus be at
least partly due to the intrinsically small number of cluster
galaxies, although it is still highly uncertain how many of the four
galaxies are in the background and how many cluster galaxies may have
been missed by the color selection: complete spectroscopic surveys or
more accurate photometric redshift measurements are needed to
determine the precise number of cluster galaxies at faint magnitude
levels.

\subsection{Properties of the Color-Selected Galaxies}

 Using the criteria described above, we investigated the properties of
the `color-selected' cluster galaxy candidates. First, we applied a
quantitative morphological classification to those galaxies in the
`$z=1.2$' color domain, using the WFPC2 $R_{\rm F702W}$ image. The
classification was performed using the central concentration index,
$C$, and the asymmetry index, $A$, to separate early-type (bulge
dominated), late-type (disk dominated), and irregular galaxies
(Abraham et al. 1996). Our procedure for measuring the $C$ and $A$ indices
followed Abraham et al. (1996), except that we adopted the centering
algorithm of Conselice et al. (1999).  Early-type galaxies show a
strong central concentration, while late-type galaxies have a lower
concentration. Irregular galaxies have a high asymmetry.

 We divided the galaxies in the `$z=1.2$' color domain into several
$R_{\rm F702W}$ magnitude bins. Since the $C$ and $A$ indices depend on
object brightness, even if the intrinsic light profiles are identical,
the boundaries between morphological types in the log $C$--log $A$
plane should be determined separately for each magnitude bin. The
boundaries are determined with the help of simulated galaxy images
produced with the IRAF ARTDATA package. For this purpose, we
constructed artificial galaxies with pure de Vaucouleurs (bulge) or
pure exponential (disk) profiles with a similar range of half-light
radii as the observed galaxies. The $C$ and $A$ indices of these
artificial galaxies were measured in an identical manner to the real
ones. As a result of this simulation, we divided the morphological
classes as shown in figure 6. Although the artificial galaxies have no
intrinsic irregularity, their apparent asymmetry arises from the
addition of noise, we adopted the 90\% completeness limit as the
boundary of the irregular galaxies in each magnitude bin. We set the
boundary between the bulge and disk-dominated galaxies so that the
fractions of correctly classified galaxies in each sample would be equal.
The resultant completeness of this bulge/disk classification is $\sim
90\%$ for the $R_{\rm F702W}$ = 24--25 mag bin and $\sim 70\%$ for the
27--28 mag bin. The resultant classification for the observed galaxies
within the `$z=1.2$' color domain is shown in figure 7.
 
 In figure 8, we show the F702W-band montage of those galaxies
having `$z=1.2$' colors. The top three rows are for those galaxies
classified as `bulge-dominated', the middle three rows are the
`disk-dominated' galaxies, and the bottom three rows are the
`irregular' galaxies. Our morphological classification seems to work
well. Each of the three rows for each morphological type represents a
range of $R_{\rm F702W}-K'$ color. It can be seen that about half of
the bulge-dominated galaxies have relatively blue colors
($R_{F702W}-K' < 5.5$).
On the other hand, there are several red ($R_{\rm F702W}-K' > 5.5$)
disk-dominated galaxies; this may imply that star-formation activity has
ceased in these objects, or it may be an effect of dust reddening,
although there may also be some contamination by `bulge-dominated' or
background galaxies. The relatively large number of `irregular'
galaxies may indicate that galaxy interactions occur frequently in the
3C 324 clusters.

 Dickinson (1997b) presented a montage of the spectroscopically
confirmed cluster members at $z =$ 1.15 and $z =$ 1.21 using the same
$R_{\rm F702W}$-band image. Although he did not identify 
things such as the coordinates, we visually compared our reduced
image with his figure 3 and confirmed that not only some red galaxies,
but also some blue galaxies in figure 8 are really cluster members. Of
the fifteen spectroscopically confirmed cluster members (excluding 3C
324 itself) in figure 3 of Dickinson (1997b), we identified seven red
($R-K > 5.5$) galaxies and three bluer galaxies within the `$z=1.2$'
color domain from their morphology and environment. We also found that
three spectroscopically confirmed cluster galaxies lie in the
foreground color domain near to the boundary ($B_{\rm F450W}-R_{\rm
F702W} \sim
-0.6$, $R_{\rm F702W}-K' \sim 3.8$). It may not be surprising that some
blue-cluster galaxies were missed in our color selection, since the
boundary of the color domain is for the ideal model galaxies and there
may be an effect of dust reddening in the UV flux.

 Figure 9 shows the C--M diagram for the galaxies in the `$z=1.2$'
color domain with their morphological classifications. A morphological
classification was not assigned for a few faint galaxies whose surface
brightness in the $R_{\rm F702W}$-band was too low.

 We compare the colors and magnitudes of the bulge-dominated galaxies
with the metallicity-sequence model for early-type galaxies calibrated
with the Coma cluster C--M relation (Kodama, Arimoto 1997) using the
Kodama and Arimoto population synthesis model (solid line). The model
galaxies are formed coevally at $z_{\rm f} =4.5$ and then evolve passively.
Clearly, the observed colors of the bulge-dominated galaxies are not
compatible with the model prediction, and the fainter objects are $\sim
1$ mag bluer than the model prediction. In fact, the bulge-dominated
galaxies in the `cluster' region seem to form a rather broad sequence
with a much steeper slope from $R_{\rm F702W}-K' \sim 6$ and $K' \sim 18$
to $R_{\rm F702W}-K' \sim 4.5$ and $K' \sim 21$.

 In figure 10, we show the sky distribution of the galaxies in the `$z
= 1.2$' color domain. The distribution of the galaxies seems to have
an irregular structure and some spatial segregation between
bright and faint sample can be seen; the redder and brighter galaxies tend
to be located around 3C 324, while the bluer or fainter galaxies have a
more diffuse distribution. While the surface densities of the
color-selected galaxies with $K < 20$ are 12.6 arcmin$^{-2}$ and 2.1
arcmin$^{-2}$ in the cluster region and the outer region,
respectively, the densities of the fainter galaxies are 7.5
arcmin$^{-2}$ and 5.9 arcmin$^{-2}$. Similarly, while the surface
densities for the red ($R_{\rm F702W}-K' > 5.5$) galaxies are 6.7
arcmin$^{-2}$ and 2.1 arcmin$^{-2}$, those for the bluer galaxies are
12.6 arcmin$^{-2}$ and 6.4 arcmin$^{-2}$. The color segregation
therefore seems to be weaker than the luminosity segregation.

\section{Discussion}

 We have presented the color distribution of the $K'$-band selected
galaxies in the region of the clusters at $z=1.2$ near to the radio
galaxy 3C 324. While the `red finger', a sequence of galaxies with
$R_{\rm F702W}-K' \sim$ 6 and $K' \sim$ 18--19 whose colors are consistent
with ages $\gtsim 3$ Gyr (or $z_{\rm f} > 4.5$; see Kodama et al. 1998) was
recognized in the color-magnitude diagram, we also found that this
sequence is truncated at $K' \sim 20$ mag; also, few galaxies in the
`cluster' region with fainter magnitudes have similar red colors.
Garilli et al. (1996) presented optical C--M diagrams ($g-r$ vs $r$)
for 67 low-redshift Abell and EMSS clusters, where most of these
clusters show a red sequence extending over more than 4 magnitudes. De
Propris et al. (1998) presented a C--M diagram ($B-R$ vs $H$) for the
Coma cluster, where the C--M relation for early type galaxies ranges
over more than 5 mag. Compared with these low-redshift clusters,
the `red sequence' of the 3C 324 clusters is clearly truncated.

 We investigated the significance of this truncation, assuming that the
red sequence population has a $K$-band luminosity distribution whose shape
is identical to the cluster population as a whole. We assumed a
faint-end slope of $\alpha = -0.9$, and adopted a characteristic magnitude
of $K^*=18.4$ (Paper I). The normalization was determined from the
observed number of red sequence galaxies with $17 < K < 20$, and the
expected number of red galaxies with $20 < K < 21$ was estimated,
taking incompleteness into account. The estimated number of red
galaxies with $5.7 < R_{\rm F702W}-K' < 6.3$ and $20 < K <21$ is 3.63 in
the cluster region and 4.85 for the entire region, while the observed
numbers are 0 and 1, respectively.
  
 We also investigated the $B_{\rm F450W}-R_{\rm F702W}$ vs $R_{\rm F702W}-K'$
two-color diagram and compared the observed colors with models of
various star-formation histories, assuming that the galaxies are
indeed at $z=1.2$. We found that the large scatter in $R_{\rm F702W}-K'$
is due to variations in both the age and star-formation activity, and
that the
objects with bluer $R_{\rm F702W}-K'$ colors have younger ages than the
`red finger' objects, provided that they are not dominated by
foreground/background galaxies.

 For moderately red objects with $4.5 < R_{\rm F702W}-K' < 5.5$ and
$B_{\rm F450W}-R_{\rm F702W} > 0$, the bluer $R_{\rm F702W}-K'$ color
is found to
be mainly due to younger ages. There is a trend that galaxies with
fainter $K'$ mag have bluer colors. Because the near-infrared
$K'$ magnitude can be an approximate measure of the total stellar mass
for quiescent galaxies, it may imply that the relatively low-mass
galaxies are, on average, younger than the massive galaxies in the 3C
324 clusters.

 The `bulge-dominated' galaxies within the cluster region seem to form
a broad color-magnitude sequence whose slope is much steeper than
predicted by coeval passive-evolution models. From the above
discussion, this can be interpreted as an effect of the age sequence
in addition to the metallicity sequence, provided that the galaxies
really are cluster members. For a comparison, in figure 9, we also plot the
`metallicity plus age' sequence model, using the Kodama and Arimoto
code. In this model, the zero point is the same as the pure
metallicity-sequence model discussed in the previous section at
$R_{\rm F702W}-K'$ = 6 ($z_{\rm f} = 4.5$, $\sim 3$ Gyr old), and the fainter
galaxies are 0.5 Gyr younger per 1 $M_{V}$-magnitude (at
$z=0.02$). The mean metallicity at each magnitude is adjusted so that
the model sequence coincides with the Coma cluster C--M relation, when
it is evolved passively to $z = 0.02$. It can be seen that this model
well reproduces the sequence of the `bulge-dominated' cluster-region
galaxies. The age difference may thus be as much as $\sim 2$ Gyr
between galaxies with $K'=18$ and $K'=21$.
 
 Is this picture consistent with the well-defined C--M sequence
observed in intermediate-redshift clusters? The C--M relation of
morphologically selected early-type galaxies in clusters at $z
=$ 0.2--0.8 is known to extend over a range of at least 3--4 magnitudes
with a scatter smaller than 0.1 mag (e.g., Ellis et al. 1997; Stanford
et al. 1998). This is very different from the C--M sequence observed at
$z=1.2$.  However, if the relatively large scatter and steep slope of
the C--M relation for the bulge-dominated galaxies in the 3C 324
clusters are caused by an age difference and the evolution is passive after
$z=1.2$, the younger (bluer) galaxies would change their color and
luminosity more rapidly than the old (red) galaxies, and may
become as red as and even fainter within a relatively short time. The
C--M relation would then become tighter and would extend over a larger
magnitude range, with a shallower slope. Note that the effect of an
additional small amount of star formation at $z = 1.2$ with a standard
IMF is negligible at intermediate-z, if the star formation ceases
immediately after $z =$ 1.2, because the brightness of these component
declines rapidly as massive stars die and the main sequence burns
down.

 For example, we compared the color and luminosity evolution of the 1
Gyr burst models with different ages, using the GISSEL96 code. One
model is 3 Gyr old at $z = 1.2$ (formation redshift $z_{\rm f} \sim 5$,
$H_{0} =$ 50 km s$^{-1}$ Mpc$^{-1}$, $q_{0}$ = 0.5) and its
optical--NIR color is $R_{\rm F702W} - K' \sim 6$, which roughly
corresponds to the `red sequence' population. The other is 1.5 Gyr old
at $z = 1.2$ ($z_{\rm f} \sim 2$) and its $R_{\rm F702W} - K' \sim 5$, which
represents the bluer galaxies. At $z \sim 0.7$, the difference in
$R_{\rm F702W} - K'$ color between these two models is only $\Delta
(R_{\rm F702W}-K') \sim 0.1$, and the difference in $B_{\rm F450W} -
R_{\rm F702W}$ color, which is more sensitive to a small amount of star
formation, becomes $\sim 0.1 $ by $z \sim$ 0.5. The fading of the
younger model galaxy in the $K'$ band is about 0.5 mag greater than
that of the old model galaxy at $z \ltsim$ 0.9; if we assume that the
younger galaxy is about 1 magnitude fainter at $z = 1.2$, it becomes
about 1.5 magnitude fainter at an intermediate redshift. From these
results, the bluer `bulge-dominated' population in the 3C 324 clusters
is expected to form a fainter part of the tight C--M relation at
intermediate-redshift, which may at least partially explain the
observed apparent truncation of the red sequence at $z=1.2$.

 In Paper I, we showed that the excess of the surface density in
the cluster region drops abruptly at $K \sim$ 20 mag. One possible
interpretation is that there may be an intrinsic deficit of faint
galaxies in the clusters. Despite the existence of color selected
galaxies with $K > 20$ mag in the cluster region, this possibility
cannot be completely ruled out, since the number of these objects may
not be large enough (subsection 3.2). There is, however, another possible
explanation, that there is a difference in the spatial distribution
between bright and faint galaxies, and that the apparent deficit of the
surface-density excess is due to the more uniform distribution of the
faint cluster galaxies. Indeed, such luminosity segregation is also
seen in the spatial distribution of the color-selected galaxies that
may be at $z \gtsim 1.2$, as shown in figure 10 (subsection 3.3). This
tendency is insensitive to subtle changes of the boundary between the
`$z=1.2$' and `foreground' color domains in figure 4. There are
several possible origins for such a luminosity segregation. Within the
framework of the theory of biased galaxy formation with cold dark
matter, luminosity segregation is naturally predicted (Valls-Gabaud et
al. 1989). In biased CDM theory, the most massive and oldest galaxies
correspond to the highest peaks of the initial density fluctuations, and
are concentrated more strongly at the early formation epoch.
Alternatively, merger events may play important roles (e.g.,
Fusco-Femiano, Menci 1998); since galaxy merging may occur more
frequently at the dense cluster core, massive galaxies can be formed
efficiently in the central region.

 The difference in the sky distribution between those galaxies with
different colors and magnitudes may be the result of the superposition
of two systems, namely those at $z =$ 1.21 and 1.15, that may
contain different galaxy populations. Postman et al. (1998) indeed
show such an example.  The cluster Cl 0023+0423 at $z =$ 0.84 was
revealed to consist of two poor clusters or groups of galaxies; they
found that there are few luminous galaxies in one system, while
luminous red galaxies exist in the other system. The 3C 324 clusters
may be in similar situation to Cl 0023+0423. Although they are not
associated with each other in the real space in the case of the 3C324
clusters, there may be a large variety of galaxy populations among the
high-redshift clusters or groups.   

 Finally, we note a rather speculative feature about the galaxy
spatial distribution. The large-scale distribution of those galaxies
with a `$z =$ 1.2' color domain in the field extends in an east--west
direction centered at 3C 324, and is aligned with the radio axis of 3C
324. This may imply a possible relationship between the past radio
jet activity of 3C 324 and galaxy formation (e.g., West 1994).\\
\vspace{0.5cm}

%*********** ACKNOWLEDGES
We wish to thank Alfonso Arag\'on-Salamanca, the referee of this
paper, for his invaluable comments. 
The present result is indebted to all the members of the Subaru
Observatory, NAOJ, Japan. We thank Nobuo Arimoto for kindly providing
the Kodama and Arimoto evolutionary synthesis models. This research
was supported by grants-in-aid for scientific research of the
Ministry of Education, Science, Sports and Culture (08740181,
09740168). This work was also supported by the Foundation for the
Promotion of Astronomy. This work is based in part on
observations with the NASA/ESA Hubble Space Telescope, obtained from
the data archive at the Space Telescope Science Institute, which is
operated by AURA, Inc.\ under NASA contract NAS5--26555. The Image
Reduction and Analysis Facility (IRAF) used in this paper is
distributed by National Optical Astronomy Observatories, operated by
the Association of Universities for Research in Astronomy, Inc., under
contact to the National Science Foundation.

%\clearpage

\section*{References}
\small

\re
Abraham R. G., van Den Bergh S., Glazebrook K., Ellis R. S., Santiago
B. X., Surma P., Griffiths R. E. 1996, ApJS 107, 1

\re
Arag\'on-Salamanca A., Ellis R. S., Couch W. J., Carter D. 1993, MNRAS 262, 764

\re
Ben\'\i tez N., Broadhurst T., Rosati P., Courbin F., Squires G.,
Lidman C., Magain P. 1999, ApJ 527, 31

\re
Bertin E., Arnouts S. 1996, A\&AS 117, 393 

\re
Bower R. G., Kodama T., Terlevich A. 1998, MNRAS 299, 1193

\re
Bower R. G.,  Lucey J. R., Ellis R. S. 1992, MNRAS 254, 601

\re
Bruzual A. G., Charlot S.  1993, ApJ 405, 538

\re
Butcher H., Oemler A. Jr 1978, ApJ 219, 18 

\re
Butcher H., Oemler A. Jr 1984, ApJ 285, 426 

\re
Cardelli J. A., Clayton G. C., Mathis J. S. 1989, ApJ 345, 245
\re
Conselice C. J., Bershady M. A., Jangren A. 1999, ApJ in press ({\it
astro-ph}/9907399)

\re
De Propris R., Eisenhardt P. R., Stanford S. A., Dickinson M. 1998,
ApJ 503, L45

\re
De Propris R., Stanford S. A., Eisenhardt P. R., Dickinson M., Elston
R. 1999, AJ 118, 719

\re
Dickinson, M. 1995, ASP Conf. Ser. 86, 283

\re
Dickinson M. 1997a, in HST and the High Redshift Universe, ed
N. R. Tanvir, A. Arag\'on-Salamanca, J.V. Wall (World
Scientific, Singapore) p.207

\re
Dickinson, M. 1997b, in The Early Universe with the VLT, ed J. Bergeron 
(Springer, Berlin) p274

\re
Dressler A., Gunn J. E. 1992, ApJS 78, 1 

\re
Dressler A., Smail I., Poggianti B. M., Butcher H., Couch W. J., Ellis
R. S., Oemler A. Jr 1999, ApJS 122, 51

\re
Ellis R. S., Smail I., Dressler A., Couch W. J., Oemler A. Jr,
Butcher H., Sharples R. M. 1997, ApJ 483, 582

\re
Fusco-Femiano R., Menci N. 1998, ApJ 498, 95

\re
Garilli B., Bottini D., Maccagni D., Carrasco L., Recillas E. 1996,
ApJS 105, 191

\re
Gladders M. D., L\'opez-Cruz O., Yee H. K. C., Kodama T. 1998, ApJ 501,
571

\re
Hall P. B., Green R. F. 1998, ApJ 507, 558

\re
HST Data Handbook, Ver3.1 1998, http://www.stsci.edu/documents/data-handbook.html

\re
Kajisawa M., Yamada T. 1999, PASJ 51, 719

\re
Kajisawa M., Yamada T., Tanaka I., Maihara T., Iwamuro F., Terada H.,
Goto M., Motohara K. et al. 2000, PASJ in press (Paper I)

\re
Kodama T., Arimoto N. 1997, A\&A 320, 41

\re
Kodama T., Arimoto N., Barger A. J., Arag\'on-Salamanca A. 1998, A\&A
334, 99

\re
Kristian J., 
Sandage A., Katem B. 1974, ApJ 191, 43 

\re
Motohara K., Maihara T., Iwamuro F., Oya S., Imanishi M., Terada H., 
Goto M., Iwai J., et al. 1998,  Proc. SPIE 3354, 659 

\re
Poggianti B. M., Smail I., Dressler A., Couch W. J., Barger A. J., Butcher H., 
Ellis R. S., Oemler A. Jr 1999, ApJ 518, 576

\re
Postman M., Lubin L. M., Oke, J. B. 1998, AJ 116, 560

\re
Rakos K. D., Schombert J. M. 1995, ApJ 439, 47

\re
Rosati P., Stanford S. A., Eisenhardt P. R., Elston R., Spinrad H.,
Stern D., Dey A. 1999, AJ 118, 76

\re
Smail I., Dickinson M. 1995, ApJ 455, L99

\re 
Smail I.,  Edge A. C., Ellis R. S., Blandford R. D. 1998, MNRAS 293,
124

\re
Spinrad H., 
Djorgovski S. 1984, ApJ 280, L9 

\re
Stanford S. A., Eisenhardt P. R., Dickinson M. 1998, ApJ 492, 461

\re 
Stanford S. A., Elston R., Eisenhardt P. R., Spinrad H., Stern D.,
Dey A. 1997, AJ 114, 2232

\re
Tanaka I., Yamada T., Arag\'on-Salamanca A., Kodama T., Miyaji T., Ohta
K., Arimoto N. 1999, ApJ in press  ({\it astro-ph}/9907437)

\re
Terlevich A. I., Kuntschner H., Bower R. G., Caldwell N., Sharples
R. M. 1999, MNRAS in press ({\it astro-ph}/9907072)

\re
Valls-Gabaud D., Alimi J.-M., Blanchard A. 1989, Nature 341, 215

\re
van Dokkum P. G., Franx M. 1996, MNRAS 281, 985

\re
van Dokkum P. G., Franx M., Kelson D. D., Illingworth G. D. 1998, ApJ 
504, L17

\re
Visvanathan N., Sandage A. 1977, ApJ 216, 214

\re
Wainscoat R. J., Cowie L. L. 1992, AJ 103, 332

\re
West M. J. 1994, MNRAS 268, 79 

\re
Yamada T., Tanaka I., Arag\'on-Salamanca A., Kodama T., Ohta K.,
Arimoto N. 1997, ApJ 487, L125

\clearpage

\begin{figure*}[p]
 \begin{center}
   \epsfile{file=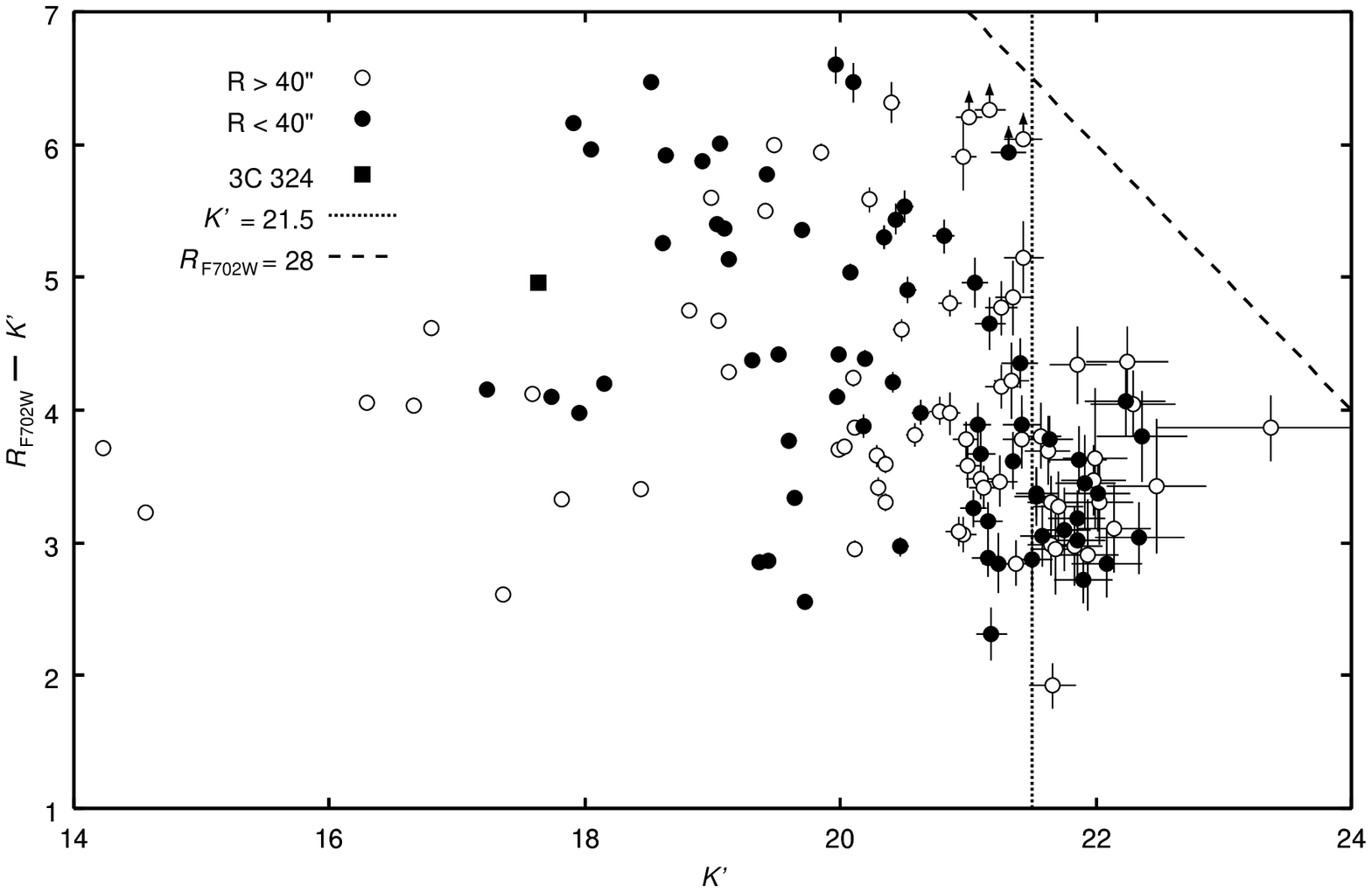,scale=0.75}
 \end{center}
\footnotesize Fig.\ 1.\ 
$R_{\rm F702W} - K'$ vs $K'$ color--magnitude diagram for the galaxies in
the region of 3C 324. The shaded and open circles represent galaxies in
the `cluster' region within 40'' of 3C 324, and the adjacent
`outer' region, respectively. 3C 324 is plotted with the solid
square. The dotted line represents $K' =$ 21.5 mag, and the dashed
line shows $R_{\rm F702W} =$ 28 mag (see text).
\end{figure*}

%\begin{fv}{2}{18pc}%
\begin{figure*}[p]
 \begin{center}
   \epsfile{file=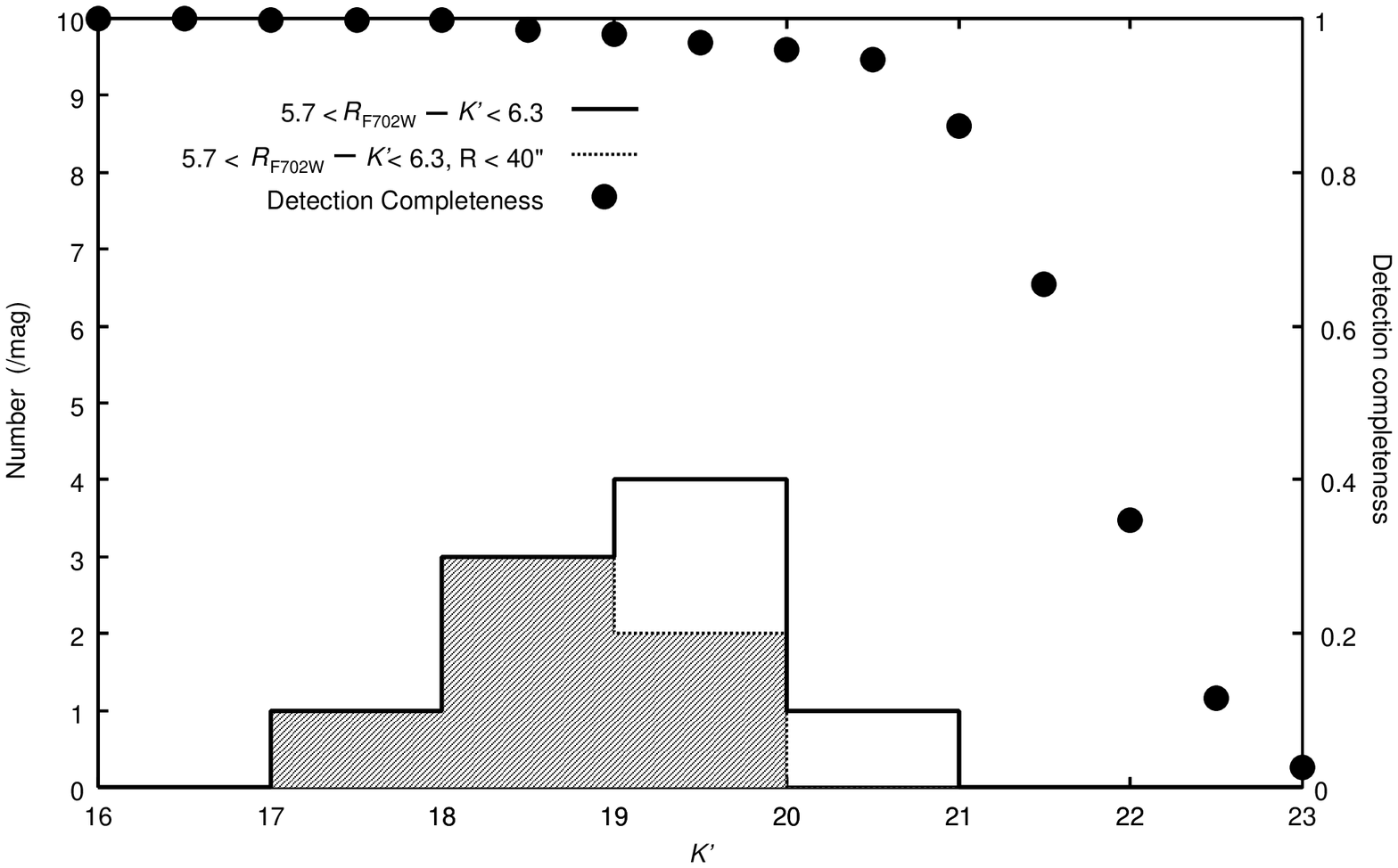,scale=0.6}
 \end{center}
\footnotesize Fig.\ 2.\ 
Number counts of the `red sequence' galaxies with 5.7 $< R_{\rm F702W} -
K' <$ 6.3, except for those objects not detected in the $R_{\rm F702W}$-band
image. The hatched region represents those galaxies within 40 arcsec
radius of 3C 324. Detection completeness in the $K'$ image is also
shown.
\end{figure*}

%\begin{fv}{3}{18pc}%
\begin{figure*}[p]
 \begin{center}
   \epsfile{file=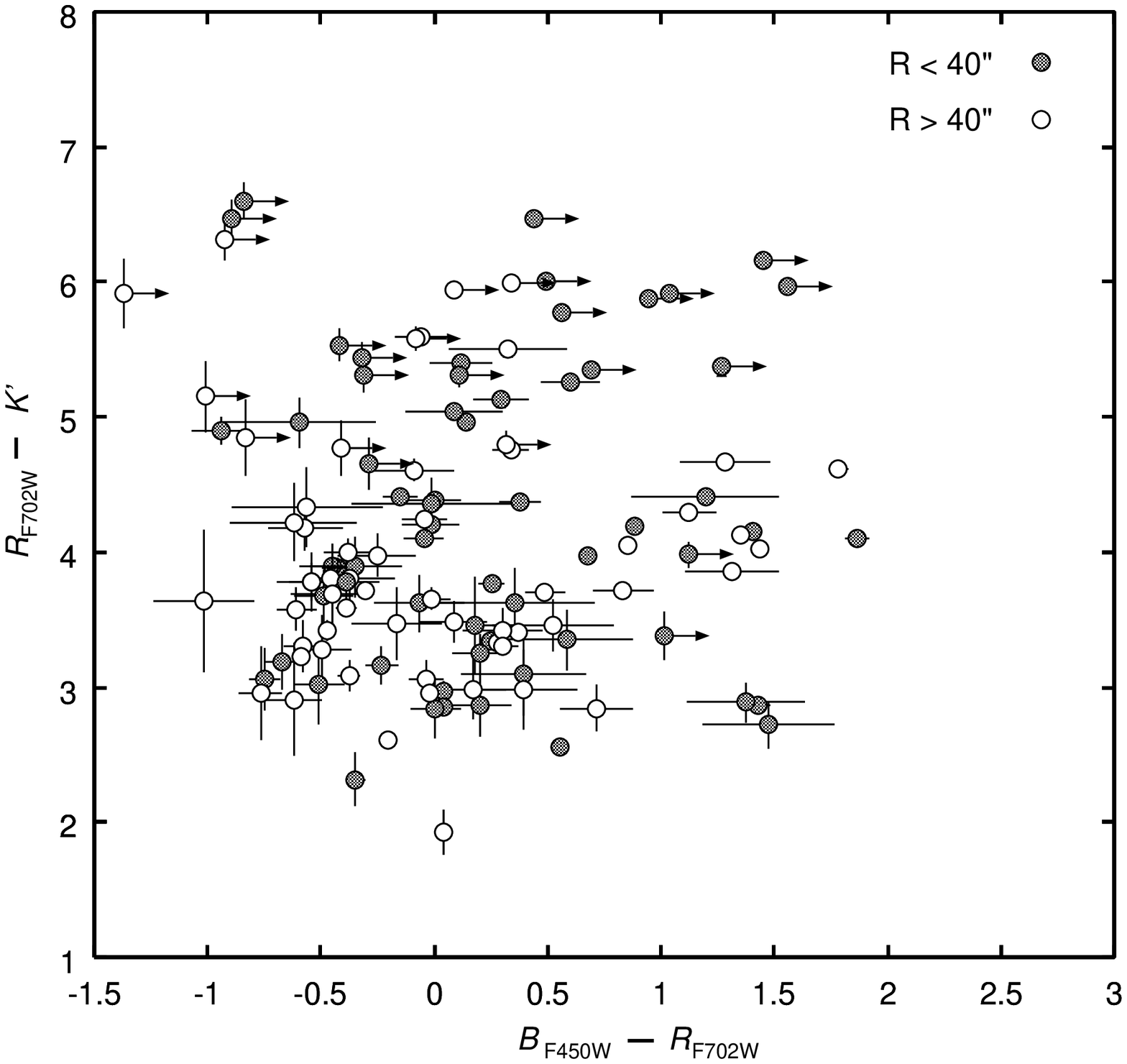,scale=0.65}
 \end{center}
\footnotesize Fig.\ 3.\ 
$R_{\rm F702W} - K'$ vs $B_{\rm F450W} - R_{\rm F702W}$ two-color
diagram for the
galaxies with $K' <$ 22. The meaning of the symbols is similar to
those in figure 1.
\end{figure*}

%\begin{fv}{4}{18pc}%
\begin{figure*}[p]
 \begin{center}
   \epsfile{file=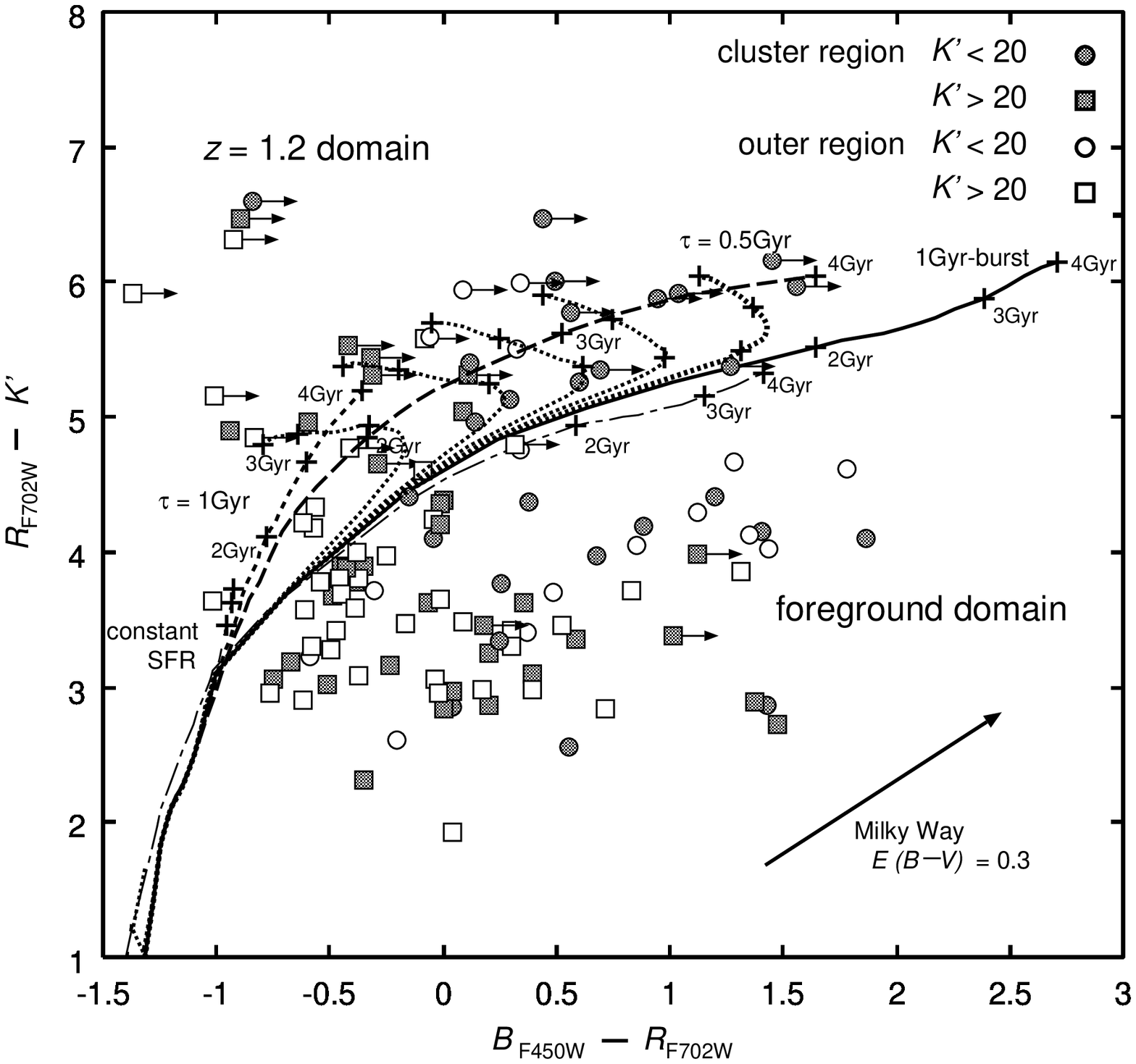,scale=0.65}
 \end{center}
\footnotesize Fig.\ 4.\ 
Same two-color diagram as figure 3, but the galaxies are divided
into brighter and fainter samples (circles for $K' <$ 20, and
squares for $K' >$ 20), and the models for the galaxies at $z = 1.2$
with various star-formation histories are plotted along the age
sequence. Tracks of the models of 1 Gyr-burst (solid line for solar
metallicity and dot-dashed line for 0.2 solar), $\tau =$ 0.5 Gyr
(long-dashed line), $\tau =$ 1 Gyr (short-dashed line), and constant
SFR (thin dashed line) are shown. The crosses represent the ages of 2,
3, 4 Gyr. The dotted lines represent the track of old (1 Gyr burst)
$+$ ongoing starburst (constant SFR) with mass fractions of 0.02,
0.05, 0.1, 0.2, 0.5\% (from right to left). We show the effect of
reddening by the large arrow using the galactic extinction curve given
in Cardelli et al. (1989) for $E(B-V) =0.3$.
\end{figure*}

%\begin{fv}{5}{18pc}%
\begin{figure*}[p]
 \begin{center}
   \epsfile{file=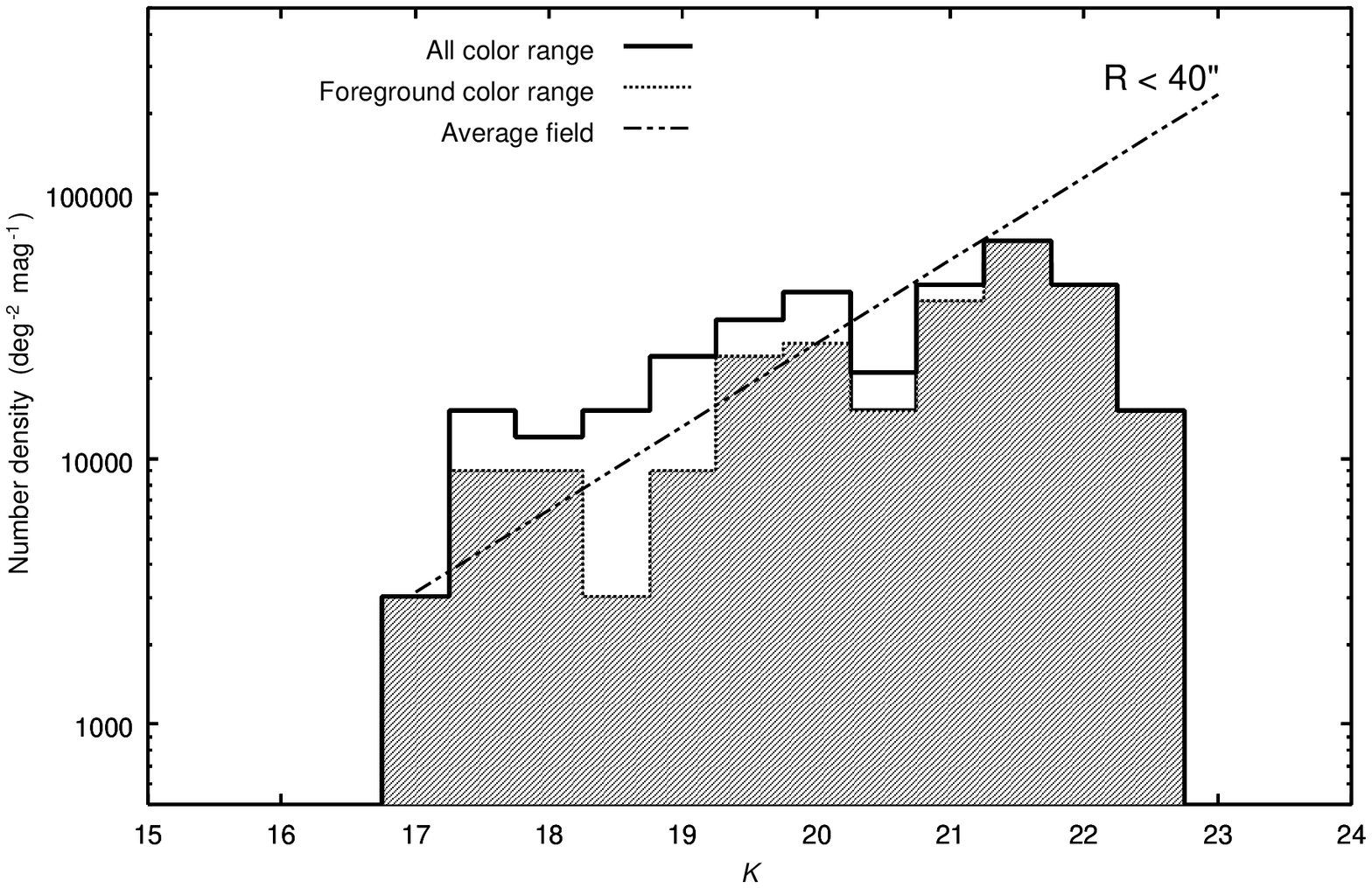,scale=0.65}
 \end{center}
\footnotesize Fig.\ 5.\ 
$K$-band number counts of galaxies in the `cluster' region. The hatched
region represents those of the galaxies within the `foreground' color
range in figure 4. The dash-dotted line shows the averaged general field
counts derived from the literature (Paper I). No color
correction is applied in converting the $K'$-band magnitude to
$K$-band one.  
\end{figure*}

%\begin{fv}{6}{18pc}%
\begin{figure*}[p]
 \begin{center}
   \epsfile{file=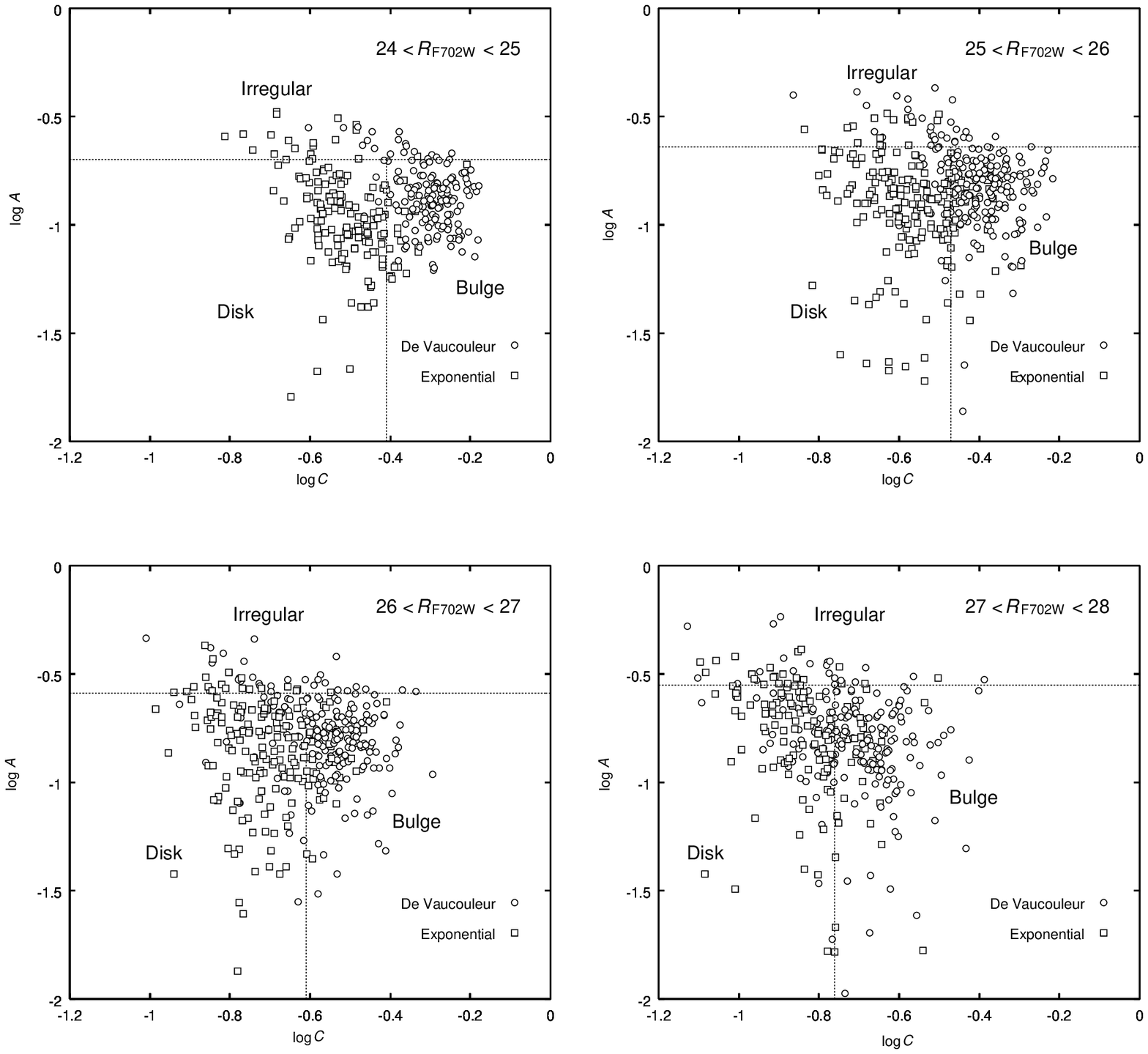,scale=0.65}
 \end{center}
\footnotesize Fig.\ 6.\ 
log $C$--log $A$ morphological classification for the artificial
galaxies on the WFPC2 image shown in each magnitude bin. Each symbol
represents the light profile of artificial galaxies. The dotted lines show
the boundary between `bulge', `disk', and `irregular' galaxies (see
text).
\end{figure*}

%\begin{fv}{7}{18pc}%
\begin{figure*}[p]
 \begin{center}
   \epsfile{file=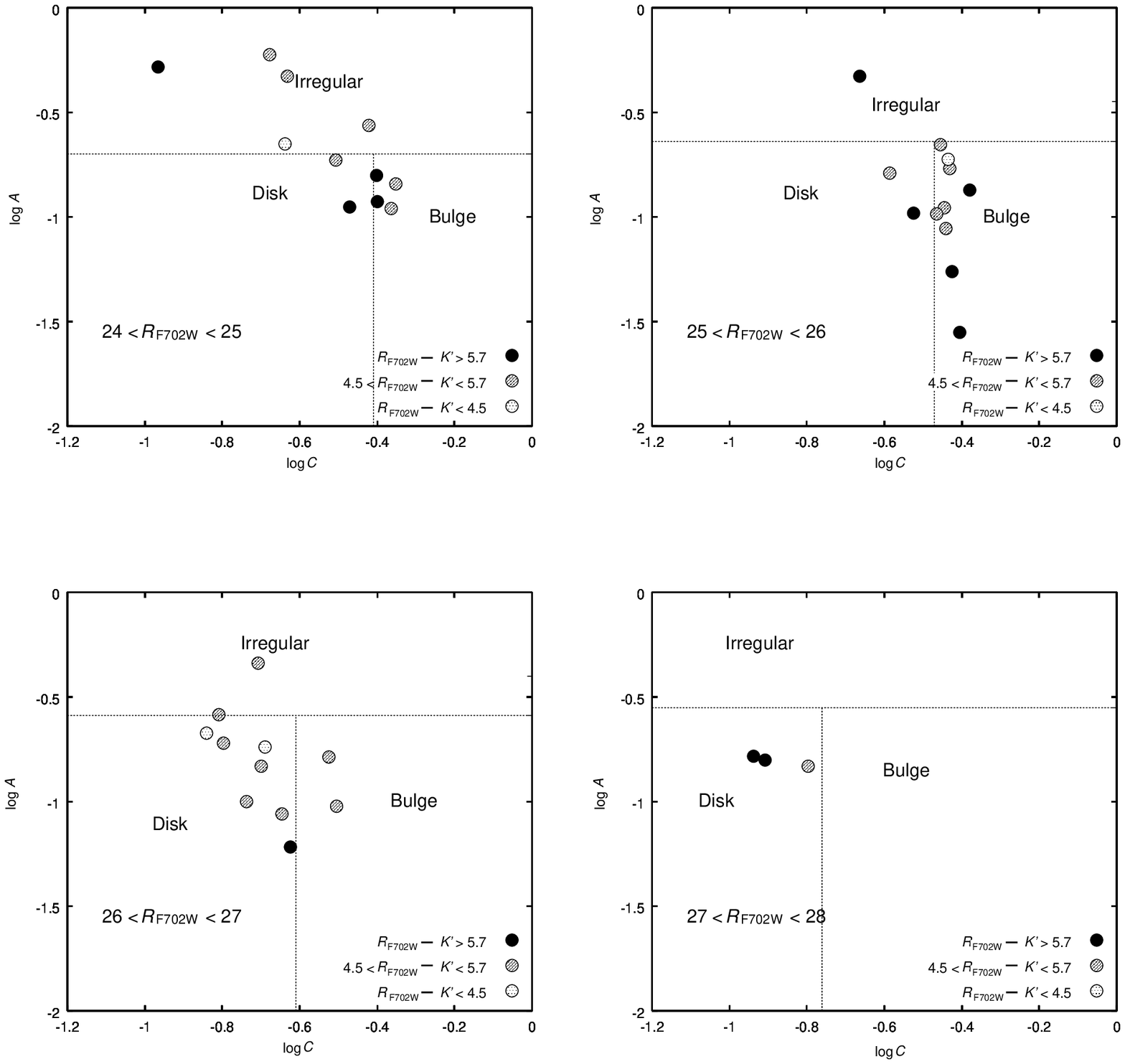,scale=0.65}
 \end{center}
\footnotesize Fig.\ 7.\ 
Morphological classification of the real galaxies within the
`$z=1.2$' color range on the WFPC2 image.
\end{figure*}

%\begin{fv}{8}{18pc}%
\begin{figure*}[p]
 \begin{center}
   \epsfile{file=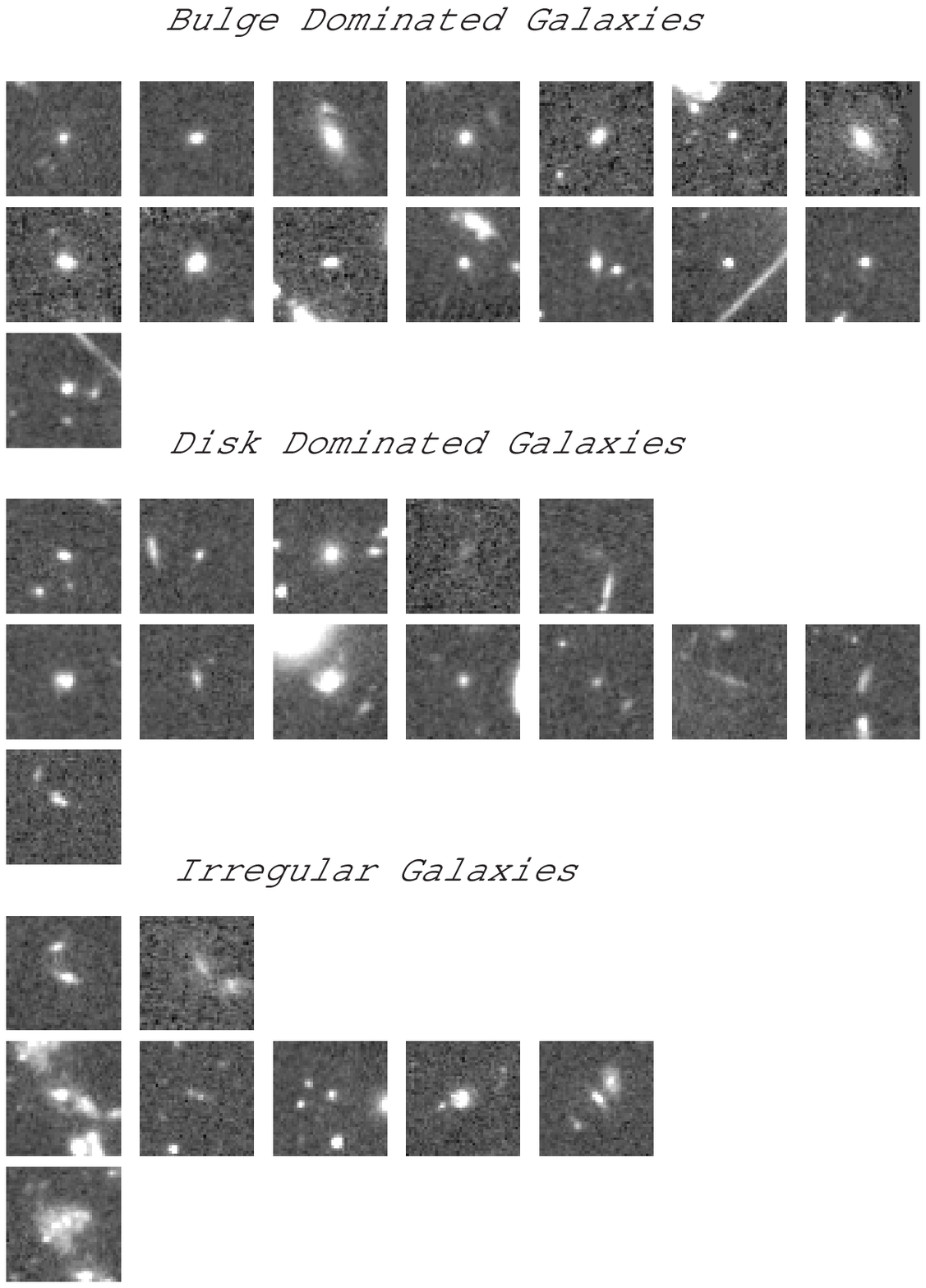,scale=1.0}
 \end{center}
\footnotesize Fig.\ 8.\ 
HST WFPC2 $R_{\rm F702W}$-band images of the galaxies in figure 7.  The top
three rows: `bulge-dominated' galaxies. Middle: `disk-dominated'
galaxies. Bottom: irregular galaxies. In each morphological block,
each row represents galaxy color; $R_{\rm F702W} - K' > 5.5$ (top), 4.5 $<
R_{\rm F702W} - K' < 5.5$ (middle), and $R_{\rm F702W} - K' <$ 4.5 (bottom).
\end{figure*}

%\begin{fv}{9}{18pc}%
\begin{figure*}[p]
 \begin{center}
   \epsfile{file=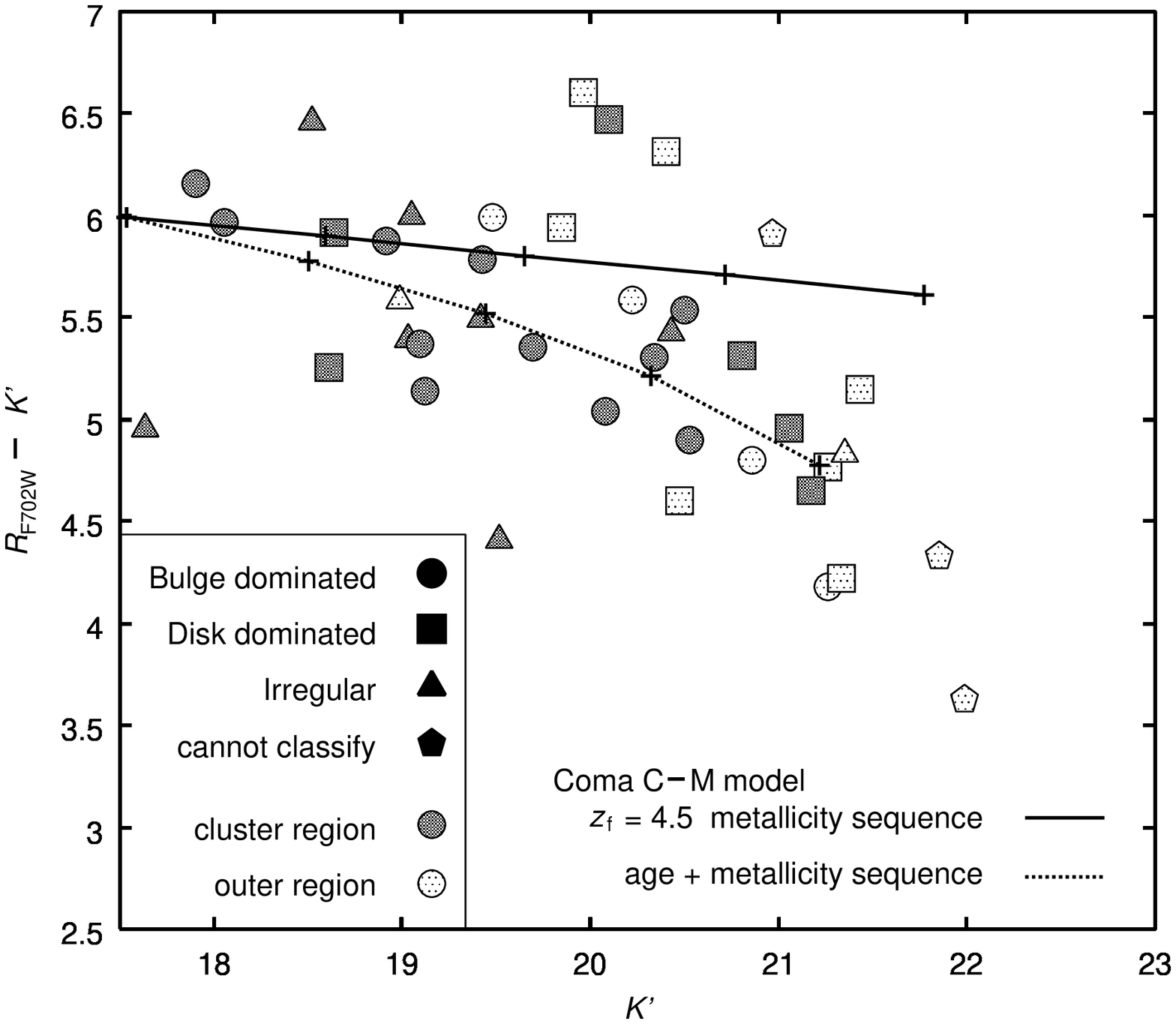,scale=0.7}
 \end{center}
\footnotesize Fig.\ 9.\ 
$R_{\rm F702W} - K'$ vs $K'$ color-magnitude diagram for those galaxies
within the `$z = 1.2$' color range in figure 4. The shape of each
symbol represents the morphology classified in figure 7. The solid
line represents the metallicity-sequence C-M relation model for the
Coma cluster (Kodama, Arimoto 1997), which would be observed at
$z=1.2$, assuming passive evolution. The dotted line shows a similar
model with a 1--3 Gyr age difference along the sequence (see text).
\end{figure*}

%\begin{fv}{10}{18pc}%
\begin{figure*}[p]
 \begin{center}
   \epsfile{file=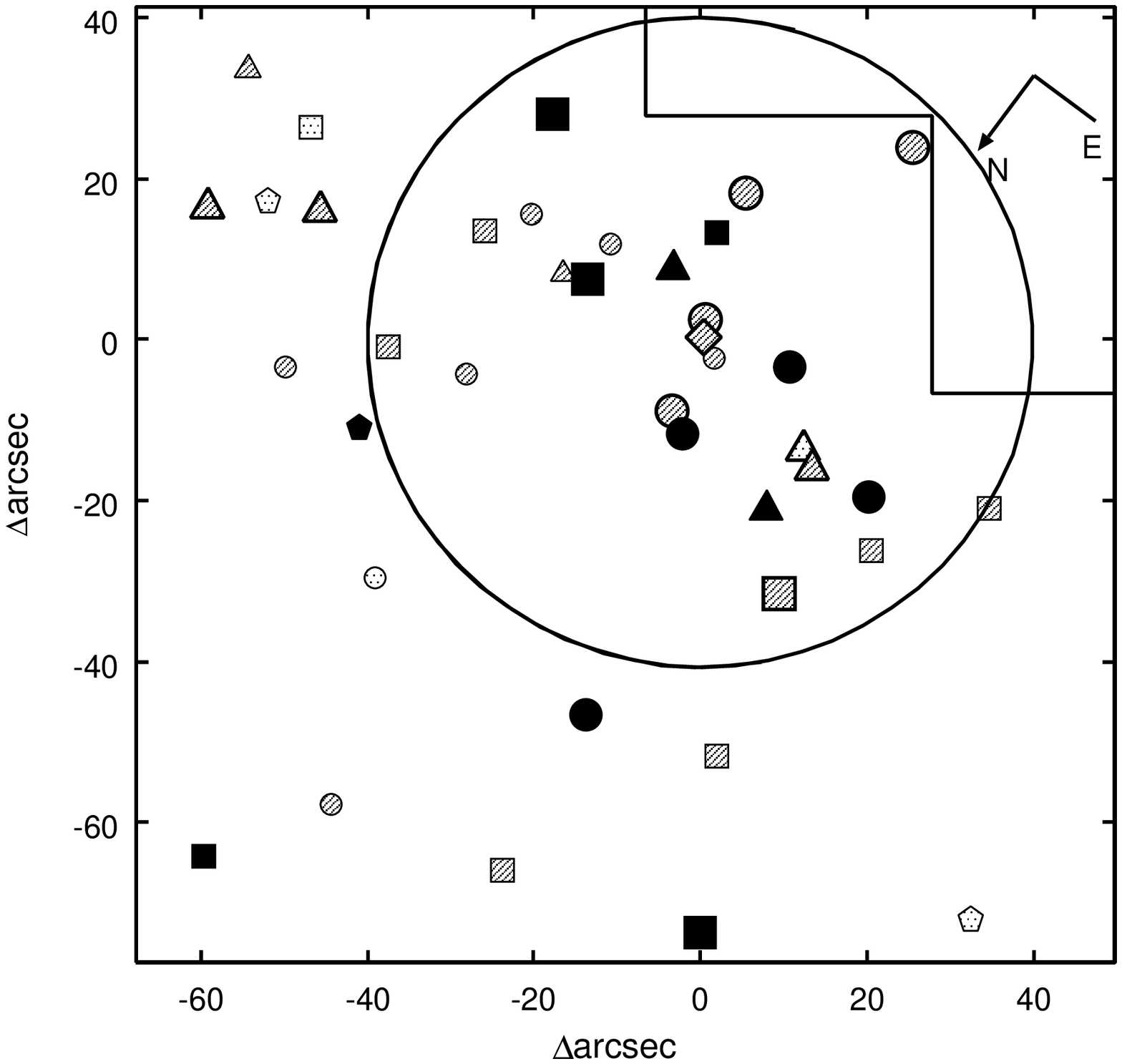,scale=0.65}
 \end{center}
\footnotesize Fig.\ 10.\ 
Spatial distribution of the galaxies within the `$z = 1.2$' color
range.  The meaning of the shape of each symbol is the same as in figure
9. The size of each symbol represents $K' < 20$ (large), $K' > 20$
(small), respectively. The pattern of each symbol represents
$R_{\rm F702W} - K'$ color, $R_{\rm F702W} - K'> 5.5$ (filled), $4.5 <
R_{\rm F702W} - K' < 5.5$ (hatched), $R_{\rm F702W} - K' < 4.5$ (dotted).  3C
324 is plotted with a diamond.  The upper light solid line shows the
boundary of the optical WFPC2 images.  The large solid circle
represents a 40-arcsec radius from 3C324.
\end{figure*}

\end{document}